\documentclass[12pt,a4paper]{article}
\pdfoutput=1
\usepackage[utf8]{inputenc}
\usepackage[T1]{fontenc}
\usepackage{jcappub}

\newcommand{\keV}{\,\text{keV}}
\newcommand{\MeV}{\,\text{MeV}}
\newcommand{\eV}{\,\text{eV}}
\newcommand{\GeV}{\,\text{GeV}}
\newcommand{\TeV}{\,\text{TeV}}
\newcommand{\cL}{\mathcal{L}}
\newcommand{\lAngle}{\langle\!\langle}
\newcommand{\rAngle}{\rangle\!\rangle}

\renewcommand{\d}{\partial}
\newcommand{\half}{\frac{1}{2}}
\newcommand{\e}{\mathrm{e}}

\title{Hiding an elephant: heavy sterile neutrino with large mixing
  angle does not contradict cosmology}
\author[a]{F. Bezrukov,}
\author[b,c]{A. Chudaykin,}
\author[b,c]{D. Gorbunov}

\affiliation[a]{The University of Manchester, School of Physics and Astronomy,\\
  Oxford Road, Manchester M13 9PL, United Kingdom}
\affiliation[b]{Institute for Nuclear Research of the Russian Academy of Sciences,\\
  60th October Anniversary prospect 7a, Moscow 117312, Russia}
\affiliation[c]{Moscow Institute of Physics and Technology,\\
  Institutsky per. 9, Dolgoprudny 141700, Russia}

\emailAdd{Fedor.Bezrukov@manchester.ac.uk}
\emailAdd{chudy@ms2.inr.ac.ru}
\emailAdd{gorby@ms2.inr.ac.ru}

\date{\today}

\abstract{We study a model of a keV-scale sterile neutrino with a
  relatively large mixing with the Standard Model sector.  Usual
  considerations predict active generation of such particles in the
  early Universe, which leads to constraints from the total Dark
  Matter density and absence of X-ray signal from sterile neutrino
  decay.  These bounds together may deem any attempt of creation of
  the keV scale sterile neutrino in the laboratory unfeasible.  We
  argue that for models with a hidden sector coupled to the sterile
  neutrino these bounds can be evaded, opening new perspectives for
  the direct studies at neutrino experiments such as Troitsk
  $\nu$-mass and KATRIN. We estimate the generation of sterile
  neutrinos in scenarios with the hidden sector dynamics keeping the
  sterile neutrinos either massless or superheavy in the early
  Universe. In both cases the generation by oscillations from active
  neutrinos in plasma is suppressed.}

\begin{document}

\maketitle
\flushbottom

\section{Introduction}

Sterile neutrinos appear in the vast majority of extensions of the
Standard Model (SM), see \cite{Abazajian:2012ys} for a review.  In
this article we concentrate on intermediate, 1--100\keV, mass range of
the sterile neutrinos \cite{Adhikari:2016bei}.  It is usually assumed
that such neutrinos have an extremely small mixing angle with the
active neutrinos, $\sin^2(2\theta)\ll10^{-7}$.  If this is not the
case, multiple astrophysical considerations contradict the existence
of such particles.  The main problem is the production in the early
Universe through the mixing with active neutrinos
\cite{Dodelson:1993je,Dolgov:2002wy}, leading to over-closure of the
Universe.  At the same time, direct laboratory bounds
\cite{Hiddemann:1995ce,Holzschuh:1999vy,Abdurashitov:2017kka} on the
active-sterile mixing in this mass range is reaching at most the
mixing angles $\sin^2(2\theta)\gtrsim10^{-3}$.

We argue that in the models with additional hidden sector coupled to
sterile neutrinos, the their generation in the early Universe can be
arbitrarily suppressed. The key ingredient is a special dynamics of
the hidden sector which changes sterile neutrino parameters
responsible for their production.  In the early Universe the most
active production of sterile neutrinos of mass $M$ by oscillations of
SM neutrinos happens at plasma temperatures
\cite{Dodelson:1993je,Abazajian:2001nj}
\begin{equation}
  \label{eq:Tprod}
  T_{\mathrm{max}} \sim 133\MeV\left(\frac{M}{1\keV}\right)^{1/3}.
\end{equation}
Thus, if at this temperature the oscillations do not
happen,\footnote{Possibly the most radical but rather constrained solution
  is to start the Hot Big Bang stage of the early Universe evolution with
  temperatures (much) below \eqref{eq:Tprod},
  see e.g.\ ref.\ \cite{Gelmini:2004ah}.} the final
abundance of sterile neutrino will be suppressed.  In this letter we
suggest two mechanisms to suppress oscillations at temperatures above
some critical temperature $T_c\ll T_{\mathrm{max}}$.

In section~\ref{sec:Generation} we outline the calculation of the
sterile neutrino abundance and velocity distribution when oscillations
are active only for $T\lesssim T_c$.  Then in section~\ref{sec:hidden}
we consider how to suppress the sterile neutrino production. We
concentrate on the natural idea that the suppression of oscillations
can be achieved by making the mass of the sterile neutrino different
in the early Universe (cf.\ variable neutrino mass models
\cite{Fardon:2003eh,Horvat:2005ua}).
Section~\ref{sec:PhaseTransition} describes the hidden sector which
undergoes phase transition, and keeps the sterile neutrino massless at
high temperatures. On the contrary, in section~\ref{sec:FrozenScalar}
we make the neutrino very heavy at early times due to the coupling
with a very light and extremely feebly interacting scalar, which was
frozen since inflation, and started oscillations only at $T\sim T_c$.

Finally, in section~\ref{sec:Discussion} we outline the overall
compatibility of the suggested framework with astrophysical
observations.  Section \ref{sec:conclusions} contains the summary.

\section{Generation of a variable mass sterile neutrino}
\label{sec:Generation}

\begin{figure}[!htb]
  \centering
  \includegraphics[width=0.5\linewidth]{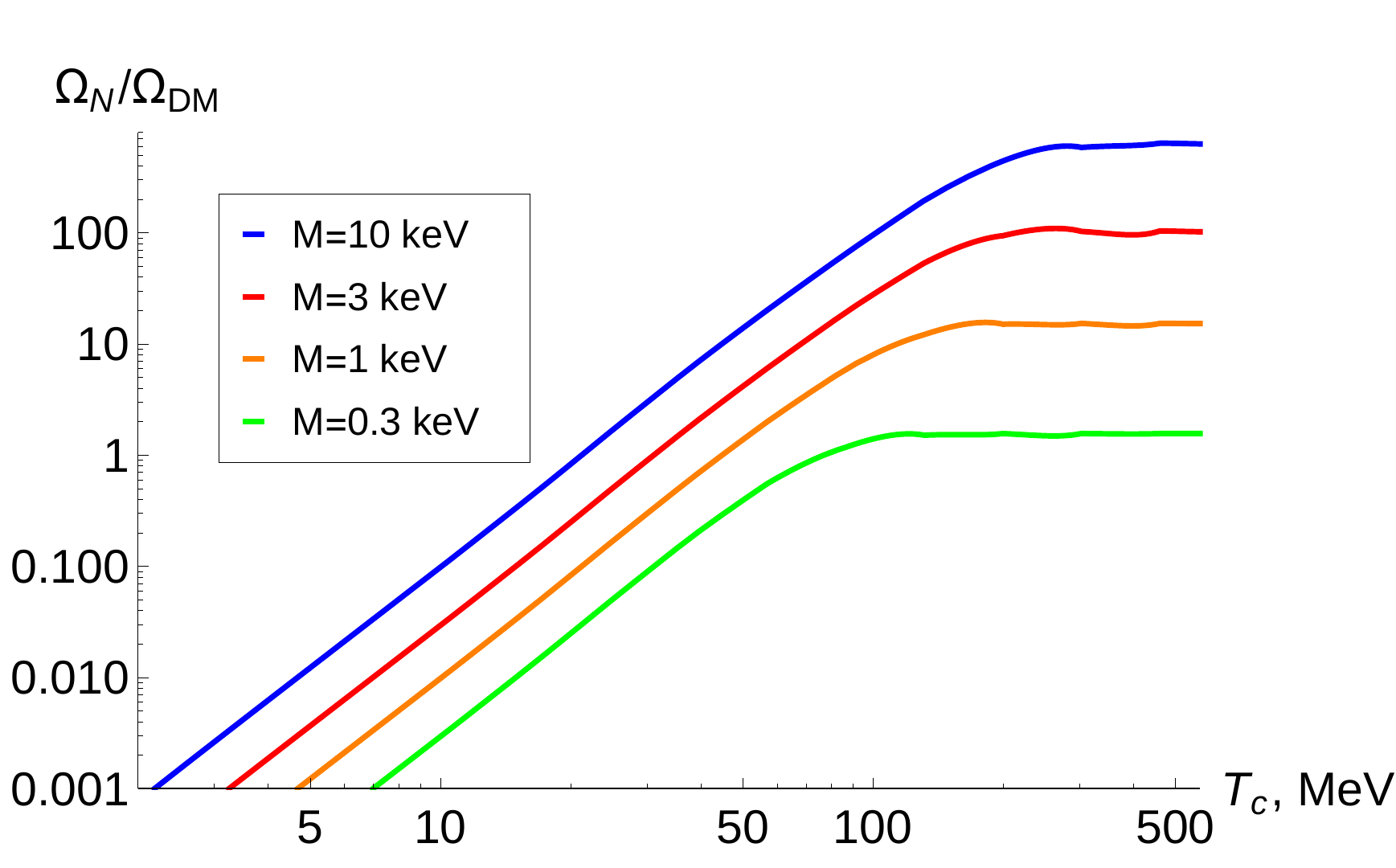}%
  \includegraphics[width=0.5\linewidth]{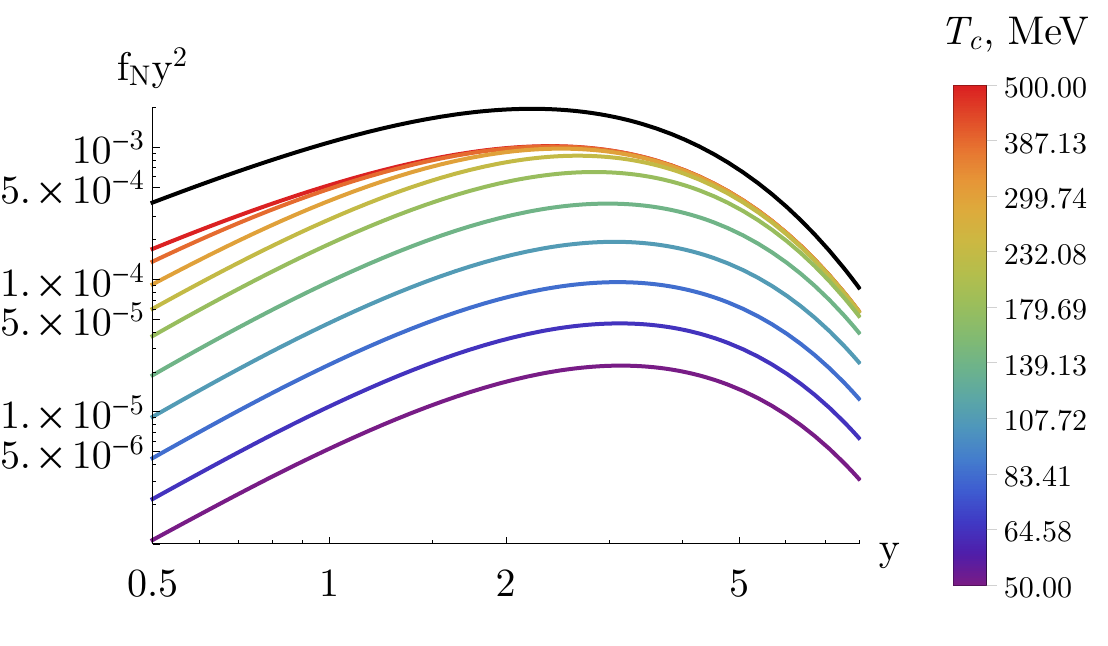}
  \caption{\emph{Left panel:} abundance of the sterile neutrino
    produced at $T<T_c$ depending on the critical temperature $T_c$
    for several masses $M$ of sterile neutrino. \emph{Right panel:}
    spectrum of the produced sterile neutrinos, with sterile neutrino
    mass $M=7.1\keV$; the top curve shows the thermal Fermi-Dirac
    spectrum for comparison.  For both panels
    $\sin^2(2\theta)=10^{-6}$ is used.}
  \label{fig:generation}
\end{figure}

The calculation of generation of sterile neutrino in the model of the
form
\begin{equation}
  \label{eq:L}
  \cL = i \bar N \hat\partial N + \frac{M}{2} \bar{N^c} N
        + y_\nu H \bar\nu_a N  + \text{h.c.},
\end{equation} 
can be made following refs.~\cite{Dodelson:1993je,Dolgov:2002wy}.  The
Yukawa term in \eqref{eq:L} gives the Dirac mass
${m_D=y_\nu\langle H\rangle}$, leading to the active neutrino mass
$m=\theta^2M$ and the active-sterile mixing angle
$\theta\simeq m_D/M$.  In the primordial plasma this mixing provides
oscillations of the active neutrinos into sterile ones. The analytic
approximation of \cite{Dodelson:1993je} breaks if number of d.o.f.\
$g_*$ changes during production, but for the most interesting region
of parameters in the present paper we examine production only at
$T\ll100\MeV$, so this approximation is completely
justified.\footnote{Strictly speaking, everywhere in the present paper
  except subsection~\ref{sec:FrozenScalar} by the temperature of the
  active sector we understand the temperature of the active neutrinos
  $T_\nu$.  With this convention we do not have to worry about changes
  of the number of d.o.f.\ after neutrino freeze-out.}  Moreover, we
assume that the number of active neutrinos is not depleted
significantly by conversion to the sterile neutrinos.  For small
mixing angles $\theta$ this is true both above and below neutrino
freeze-out temperature $T_{\nu,f}\sim2\MeV$.  With this assumptions
the distribution function $f_N(T,E/T)$ of the sterile neutrino evolves
according to the equation
\begin{equation}
  \label{eq:BoltzmanRate}
  HT \left( \frac{\partial f_N}{\partial T} \right)_{E/T} =
    \left[ \frac{1}{2}\sin^2(2\theta_M) \right] \frac{\Gamma_A}{2}f_A,
\end{equation}
where the effective angle and rate of active neutrino scattering in
matter, in assumption of mixing only between the sterile and electron
neutrino\footnote{The overall approach described here can be applied
  to the case of mixing with muon and tau neutrino as well, with the
  appropriate change of the numerical factors in $\Gamma_A$ and $c$.
  For more detailed information see e.g.~\cite{Asaka:2006nq}.}
\cite{Enqvist:1991qj}, reads
\begin{equation}
  \label{matter-mixing}
  \sin^2(2\theta_M)=\frac{m_D^2}{m_D^2+[c\Gamma_A E/M+M/2]^2},
  \qquad
  \Gamma_A\approx 1.27\times G_F^2 T^4 E,
\end{equation}
with $c\approx 63$~\cite{Asaka:2006nq}.  The solution to equation
\eqref{eq:BoltzmanRate} in the expanding Universe with constant $g_*$
is
\begin{equation}
  \label{eq:fSfA}
  \frac{f_N}{f_A} = \frac{2.9}{g_*^{1/2}}
     \left(\frac{\theta^2}{10^{-6}}\right) \left(\frac{M}{\keV}\right)
     \int_{x}^{x_c} \frac{y\,dx'}{(1+y^2x'^2)^2},
\end{equation}
where $y\equiv E/T$, $x\equiv 148(T/\GeV)^3(\keV/M)$ with $T$
corresponding to the observation moment and $T_c$ to the critical
temperature.  Here $f_A(y)=1/(\e^y+1)$ is the distribution function of
the active neutrinos, and $g_*=10.75$ is the number of d.o.f.\ just
before the active neutrino freeze-out.  At late times we can put the
lower integration limit to $x=0$, and for $T_c\ll100\MeV$ we can
expand the expression for small $x_c$, arriving at sterile neutrino
distribution
\begin{equation}
  \label{eq:fSfAapprox}
  \frac{f_N(y)}{f_A(y)} \simeq
    0.13\times\theta^2\left(\frac{10.75}{g_*}\right)^{1/2}
    \left(\frac{T_c}{\MeV}\right)^3\cdot y.
\end{equation}
This is a warm distribution with average momentum
$\lAngle p\rAngle=4.1T$, as compared to the usual thermal average
$\lAngle p\rAngle=3.1T$.  The number density for the distribution
(\ref{eq:fSfAapprox}) gives the present sterile neutrino contribution
to the matter density
\begin{equation}
  \label{eq:Omegalate}
  h^2\Omega_N \simeq
    4.3 \times\theta^2 \left(\frac{10.75}{g_*}\right)^{1/2}
    \left(\frac{T_c}{\MeV}\right)^3\left(\frac{M}{\keV}\right)
\end{equation}
which is strongly suppressed for small $T_c$.  In
figure~\ref{fig:generation} we show the results of direct integration
of (\ref{eq:BoltzmanRate}) with temperature dependent $g_*$ and
$\Gamma_A$ following refs.~\cite{Abazajian:2001nj,Ghiglieri:2015jua}
(for improved calculations at $T\sim T_{\mathrm{max}}$ see
\cite{Ghiglieri:2015jua}).  This demonstrates the applicability of
(\ref{eq:Omegalate}) for the interesting range of $T_c$.

A remarkable thing is that with oscillations suppressed at high
temperatures the sterile neutrino \emph{can be involved into the
  seesaw type I mechanism
  \cite{Minkowski:1977sc,GellMann:1980vs,Mohapatra:1979ia,Yanagida:1980xy,Schechter:1980gr}
  of giving mass to the active neutrino.} This is contrary to the
cases with usual non-resonant or resonant dark matter sterile neutrino
production (see e.g.\ ref.~\cite{Adhikari:2016bei} for
details). Indeed, typical values of the mixing angle following from
the seesaw expression of active neutrino mass $m$ are
\begin{equation}
  \label{seesaw}
  \frac{\theta^2}{10^{-4}} = \frac{1\keV}{M}\frac{m}{0.1\eV}.
\end{equation}
For active neutrino masses of order of solar
$\sqrt{\delta m_\mathrm{sol}^2}\simeq 0.9\times10^{-2}\eV$ or
atmospheric $\sqrt{\delta m_\mathrm{atm}^2}\simeq0.05\eV$ neutrino
masses and $T_c$ in MeV range, the sterile neutrino contribution to
Dark Matter \eqref{eq:Omegalate} can be small enough, $\Omega_N\ll1$,
to make astrophysical constraints irrelevant (see
section~\ref{sec:Discussion} for details).  Note, that with multiple
sterile neutrinos in the model mixing angle $\theta$ can be larger
than \eqref{seesaw} due to cancellations in the mass matrix.

\section{Possibilities with a hidden sector coupled to sterile neutrinos}
\label{sec:hidden}

To suppress the oscillations \eqref{eq:BoltzmanRate} in the early
Universe plasma one must reduce the effective mixing angle in matter
$\theta_M$~\eqref{matter-mixing}.  This can be achieved by introducing
a hidden sector coupled to the sterile neutrinos. Then two options are
straightforward: arrange the dynamics of the hidden sector and its
coupling to neutrinos in such a way that the sterile neutrinos in the
early Universe become either massless or super-heavy. The first option
may be realized with late-time phase transition in the hidden sector,
which makes the sterile neutrino massive. This option is considered in
section~\ref{sec:PhaseTransition}. The second option may be realized
by either sufficiently strong interaction of sterile neutrino with
hidden sector plasma (which cools down while the Universe expands) or
by neutrino coupling to a classical field frozen in the early Universe
and forming Bose--Einstein condensate at present. The later
realization is studied in section~\ref{sec:FrozenScalar}, while
investigation of the former is deferred to the future.

\subsection{Phase transition in the hidden sector}
\label{sec:PhaseTransition}

This scenario allows to change the Majorana mass of sterile neutrino
$M$ from zero in the early Universe (so that the sterile and active
neutrino pair forms a pure Dirac fermion) to the present day value at
keV scale.  In this scenario we assume existence of an extended hidden
sector which is connected to the active sector (SM) only via sterile
neutrinos (neutrino portal).  We also assume that the hidden sector
has its own temperature below $T_h=\xi T$, with $\xi\ll1$ (a setup
similar to the one adopted in Mirror World models, for review see
e.g.~\cite{Foot:2014mia}).  With the hidden sector temperature being
below that of the active sector we are not concerned with the hidden
sector contribution to the energy density of the Universe starting
from the generation of sterile neutrinos and until today.  We also
assume that one of the hidden scalars\footnote{If $\phi$ is a complex
  scalar rather it carries the lepton charge and becomes a Majoron,
  leading to more complicated phenomenology, see e.g.\ ref.\ \cite{Bento:2001xi}.  For the main purpose of
  our paper this is not important, and we proceed with the real
  scalar.} $\phi$ interacts with the sterile neutrino
\begin{equation}
  \label{eq:LNh}
  \cL = \frac{f}{2} \phi \bar{N^c} N + \text{h.c.},
\end{equation}
so that the Majorana mass of the sterile neutrino entering Lagrangian
\eqref{eq:L} is generated by its expectation value,
$M=f\lAngle\phi\rAngle$.

The required behaviour of the sterile neutrino mass is achieved in the
situation with phase transition happening in the hidden sector at
temperature $T_{h,c}=\xi T_c$, so that $\phi$ is in the symmetric
state at high temperatures,
\begin{equation*}
  \lAngle\phi\rAngle|_{T_h>\xi T_c} = 0,
\end{equation*}
and the symmetry is broken at low temperatures,
\begin{equation*}
  \lAngle \phi \rAngle|_{T_h<\xi T_c} = v_\phi,
  \qquad M=fv_\phi.
\end{equation*}
With an instant phase transition in the hidden sector, the spectrum of
sterile neutrinos produced at later stages of evolution is given by
eq.~\eqref{eq:fSfA}.

We find in (\ref{eq:Omegalate}) that abundance of sterile
neutrino can be strongly suppressed by decreasing the critical
temperature $T_c$.  Let us find the maximal suppression we can expect.
In eq.~(\ref{eq:fSfAapprox}) we assumed that the abundance of sterile
neutrino is exactly zero at $T_c$.  Note, that in this scenario $M$ is
zero at higher temperatures, so we can approximate neutrinos as Dirac
particles of mass $m_D$, and expect the admixture of right handed
components in the primordial plasma at the level of
$m_D^2/4p^2=m_D^2/4y^2T^2$.  Therefore, in addition to
(\ref{eq:fSfAapprox}), there is also an \emph{initial} contribution to
the sterile neutrino spectrum
\begin{equation}
  \label{eq:fSfAinitial}
  \frac{f_{S,\text{in}}}{f_A} \simeq \frac{m_D^2}{4y^2T_c^2}
    \simeq \frac{0.25\times10^{-6}\theta^2}{y^2} \left(\frac{M}{\keV}\right)^2
    \left(\frac{\MeV}{T_c}\right)^2,
\end{equation}
where we expressed $m_D^2=\theta^2 M^2$ using the present day values
of the mixing angle and Majorana mass. This effect makes the
contribution to the sterile neutrino fraction in the energy density of
the present Universe
\begin{equation}
  \label{Omega-initial}
  h^2\Omega_{N,\text{in}} =
    \frac{MT_{\nu,0}^3}{(\rho_c/h^2)}\frac{2}{2\pi^2}
    \int_0^\infty dy\,y^2 f_{S,\text{in}}(y)
  \approx 10^{-6}
    \theta^2 \left(\frac{M}{\keV}\right)^3
    \left(\frac{\MeV}{T_c}\right)^2. 
\end{equation}
Equating the estimates for the initial \eqref{Omega-initial} and late
\eqref{eq:Omegalate} contributions we get that they become equal at
critical temperature\footnote{In formula \eqref{eq:Omegalate} we used
  here $g_*\approx13$ valid for the critical temperature of the active
  neutrino sector below $1\MeV$.}
\begin{equation}
  \label{eq:Tcmin}
  T_{c,\text{min}}\simeq 0.05 \MeV \left(\frac{M}{\keV}\right)^{2/5}.
\end{equation}
Thus, the minimal sterile neutrino abundance at given mass can be
estimated as
\begin{equation}
  \label{eq:Omegamin}
  h^2\Omega_{N,\text{min}} \simeq h^2\Omega_{N,T<T_C}+h^2\Omega_{N,\text{in}}
  \simeq 0.9\times10^{-3}\theta^2\left(\frac{M}{\keV}\right)^{11/5}.
\end{equation}
The final spectrum of sterile neutrinos in the model is given by the
sum of generated \eqref{eq:fSfAapprox} and initial
\eqref{eq:fSfAinitial} contributions,
$f_N=f_{S,T<T_C}+f_{S,\text{in}}$.  The former spectrum is warm, while
the latter is cooled as compared to the thermal Fermi--Dirac spectrum.
The presence of a cool component weakens the structure formation
bounds on the DM.  However, as we will see in
section~\ref{sec:Discussion}, this effect is insufficient to allow for
generation of all of the DM by the described mechanism.

We do not provide a full and complete realization of the hidden sector
in the present note, but rather list here several requirements that
should be satisfied by the hidden sector model.  The reader is invited
to formulate a concrete realization.  First, the phase transition
should be fast enough.  Second, it is preferable to have the
interactions in the hidden sector in the weak coupling regime.  This
could put non-trivial constraints on the mass spectrum of the sector.
Third, our analysis is applicable only if the interactions due to the
Yukawa term in (\ref{eq:LNh}) do not significantly modify the
oscillation picture, that is the rate of sterile neutrino scattering
is be much smaller than $\Gamma_A$.

\subsection{Feebly interacting scalar}
\label{sec:FrozenScalar}

Let us investigate the opposite dynamics in the scalar sector: instead
of rapidly interacting sector, we use a free massive scalar which has
only a feeble Yukawa coupling (\ref{eq:LNh}) to sterile neutrino,
$f\ll1$.  Let the scalar be extremely light.  The model Lagrangian for
the dark sector contains, in addition to (\ref{eq:LNh}),
\begin{equation}
  \cL_{DS}= \half(\d \phi)^2-\half m_\phi^2 \phi^2.
\end{equation}
The radiatively induced scalar self-coupling $\sim f^4/(4\pi)^2$ is
negligibly small, and we consider again only the real field
$\phi$. Due to the Hubble friction, the scalar field remains constant
in the hot Universe, and its field value $\phi_i$ is determined by the
pre-Big-Bang history: either inflation or preheating.  In particular,
as far as the scalar is effectively massless at inflation, its value
can be more or less arbitrary at inflationary stage in the observed
Hubble patch of the Universe.

If the initial field is large, the sterile neutrinos can be very
heavy.  Let us assume the Yukawa coupling $y_\nu$ in \eqref{eq:L} is
sufficiently small to prevent the direct sterile neutrino production
in the hot plasma of the early Universe.  Then after the EW transition
the sterile neutrino production can be forbidden by kinematics, if
sterile neutrinos are heavy enough,
\[
  M_{N,i}=f\phi_i\gg T.
\]
For example, sterile neutrinos of $M_{N,i}=200\GeV$ will not be
produced. Much later, when the Hubble rate drops below the scalar
field mass,
\begin{equation}
  \label{osc-start}
  H_{\text{osc}}\simeq m_\phi,
\end{equation}
the scalar field starts to oscillate with decreasing amplitude, 
\begin{equation}
  \label{5*}
  \phi\propto a^{-3/2},
\end{equation}
and hence the scalar contribution to neutrino mass drops as well.
Thus, at late times the scalar contribution (\ref{eq:LNh}) to the
sterile neutrino mass becomes negligible, and the present day mass is
given by the bare mass term in (\ref{eq:L}).\footnote{Another option
  to have neutrino massive after the oscillation start is
  non-renormalizable interaction (instead of he Yukawa coupling)
  $f\frac{\phi^2}{\Lambda}\bar N^cN$ with an interesting option of
  gradually decreasing sterile neutrino mass including the present
  epoch; we do not study this option here, see however
  ref.~\cite{Zhao:2017wmo,Vecchi:2016lty} for the case of similar mechanism for
  active neutrino.}

With the temperature of the onset of scalar field oscillations
$T_{\text{osc}}$ being sufficiently high, but lower than
$T_{\mathrm{max}}$ \eqref{eq:Tprod}, we can face the situation when
the present day contribution of (\ref{eq:LNh}) to sterile neutrino
mass is negligible, and sterile neutrino production via oscillations
in the early Universe \eqref{eq:BoltzmanRate} is strongly suppressed
as well.

Let us estimate the minimal value of critical temperature $T_c$
leading to the described scenario.  For this, we assume that the
interaction \eqref{eq:LNh} does not contribute significantly to the
present mass, $f\phi_0\ll M$.  Together with the assumption of MD
scaling of the field \eqref{5*} during its oscillations this leads to
the inequality\footnote{Hereafter in this subsection $T$ denotes the
  photon temperature.}
\[
  \left(\frac{T_\mathrm{osc}}{100\eV}\right)
  \left(\frac{2.73\,\text{K}}{T_0}\right) 
  \gtrsim
  \left(\frac{M_{N,i}}{1\TeV}\right)^{2/3}
  \left(\frac{1\keV}{M}\right)^{2/3}.
\]
This means that the described picture breaks only for $T_\text{osc}$
below $100\eV$.  One concludes that the sterile neutrino mass can
start from TeV scale and reach keV range by the present epoch. In this
case sterile neutrinos always remain non-relativistic in plasma and
hence completely avoid production in oscillations: no contribution to
the relic abundance from eq.~\eqref{eq:Omegalate}.  Thus, the sterile
neutrino production by oscillations, which might happen in the plasma
after the electroweak phase transition can be ultimately suppressed in
this model.

As an interesting additional feature, the hidden scalar of the model
of this section can be a DM candidate.\footnote{Unless the
  oscillations begin at plasma temperature below $200\eV$, which is
  the latest time when Dark Matter must appear in the
  Universe~\cite{Sarkar:2014bca}.}  Let us consider the late-time
cosmology of the hidden sector, assuming the oscillations starting at
the radiation domination, $T_{\text{osc}}>T_\mathrm{RD/MD}$. The
energy remains in the oscillating scalar all the way until the present
time. The scalar relic in the Universe contributes to the present dark
matter component an amount of
\begin{equation}
  \label{scalar-to-DM}
  \rho_{\phi,0}=\half m_\phi^2\phi_i^2( 1+z_{\text{osc}})^{-3},
\end{equation}
where $z_*$ is redshift of the oscillation epoch
\eqref{osc-start}. The scalar field saturates the dark matter if
\begin{equation}
  \label{scalar-DM}
  \rho_{\phi,0}=\epsilon_\phi\Omega_{\text{DM}}\rho_c,
\end{equation}
with $\epsilon_\phi=1$, otherwise contributes a fraction of
$\epsilon_\phi$ to the dark matter density.  The oscillations onset is
given by the condition \eqref{osc-start}, and for
$T_\mathrm{osc}\lesssim 1\MeV$ one obtains\footnote{For higher
  oscillation onset temperature $1\MeV<T_\mathrm{osc}<T_\mathrm{max}$,
  that is for heavier scalars, the calculation should be
  straightforwardly corrected for the non-constant number of degrees
  of freedom in the primordial cosmic plasma.}
\begin{equation}
  \label{Hubble-RD}
  m_\phi^2=  H_{\text{osc}}^2 =
    H_0^2\Omega_{\text{rad}}\left(1+z_{\text{osc}}\right)^4 =
    H_0^2\Omega_{\text{rad}} \frac{T_{\text{osc}}^4}{T_0^4},
\end{equation}
with $H_0$ and $T_0$ referring to the present Hubble parameter and
relic photon temperature.  Then from eqs.~\eqref{scalar-DM},
\eqref{scalar-to-DM}, \eqref{Hubble-RD} one finds for the initial
amplitude
\[
  \phi_i^2 = \frac{6M_{\text{Pl}}^2}{8\pi}
    \frac{\Omega_\mathrm{DM}\epsilon_\phi}{\Omega_\mathrm{rad}}
    \frac{1}{1+z_{\text{osc}}},
\]
where the redshift of the oscillation onset is
\[
  1+z_{\text{osc}} =
    \frac{1}{\Omega_\mathrm{rad}^{1/4}} \left(\frac{m_\phi}{H_0}\right)^{1/2}.
\]
With oscillation temperature constrained from above by
\eqref{eq:Tprod}, $T_\mathrm{osc}\ll T_\mathrm{max}\sim 100\MeV$, the
scalar mass must be below
\[
  m_\phi \sim H_{\text{osc}} = \frac{T_\mathrm{osc}^2}{M_{Pl}^*} \ll
    \frac{T_\mathrm{max}^2}{M_{Pl}^*} \lesssim 10^{-11}\eV. 
\]
Evidently, this forbids scalar perturbative decay into keV scale
sterile neutrinos. The production by oscillating scalar at zero
neutrino mass crossing is also suppressed in the wide region of
parameter space.

Overall, for the mechanism outlined in this section the DM can be
explained by the oscillating scalar field condensate. The
active-sterile neutrino mixing can be arbitrary large as far as
cosmology and astrophysics are concerned.

\section{Discussion of cosmological and astrophysical constraints}
\label{sec:Discussion}

Sterile neutrino models are constrained by astrophysical observations
and by direct searches.  We start with summarising the present status
of astrophysical constraints and combine them with the generation
mechanisms described in the previous sections.  Later we compare the
results with the current and future potential of the direct laboratory
searches.

We consider here three sources of constraints from cosmology and
astrophysics.\footnote{Sterile neutrinos of
    masses and mixing angles in the interesting ranges may contribute
    to other astrophysical processess, like supernovae explosions, 
    Universe reionization etc, see reviews
    \cite{Dolgov:2002wy,Abazajian:2012ys,Adhikari:2016bei}.
    However, these processes are quite involved by itself and application of the
    results found in literature to our models with hidden sector
    is not straightforward.}
First constraint comes from the requirement that the sterile
neutrino contribution to the energy density does not exceed the DM
contribution, $\Omega_N\leq\Omega_\mathrm{DM}$.

Second, due to the mixing with active neutrino there is a radiative
decay mode of the sterile neutrino, $N\to\gamma\nu$, leading to the
monochromatic photons of energy $E=M/2$ arriving from the relic
sterile neutrino background.  Cosmic X-ray observations place upper
bounds on the photon flux at a given energy, which limits the
sterile-active mixing angle, see \cite{Adhikari:2016bei} for review.
Let us denote by $\theta_\text{X-ray}(M)$ the limiting value in the
assumption that sterile neutrino composes all of the DM.  Then, if the
sterile neutrino composes only fraction of the DM, the bound is
rescaled as
\begin{equation}
  \label{eq:Xray}
  \sin^22\theta (M) <
    \frac{\Omega_\mathrm{DM}}{\Omega_N}\sin^22\theta_\text{X-ray}(M).
\end{equation}

Third, if the sterile neutrinos compose a significant part of the DM,
$\Omega_N\lesssim \Omega_\mathrm{DM}$, one has to take into account
bounds from the structure formation, arriving from the relative
abundance of dwarfs and other small satellite galaxies
\cite{Bode:2000gq}, phase space density of galactic dark matter
\cite{Tremaine:1979we}, and analyses of Lyman-$\alpha$ forest
\cite{Narayanan:2000tp}. These bounds constrain the level of free
streaming in the dark matter, so they actually limit the average
velocity of the dark matter particles.

\begin{figure}[!htb]
  \centering 
  \includegraphics[width=0.75\linewidth]{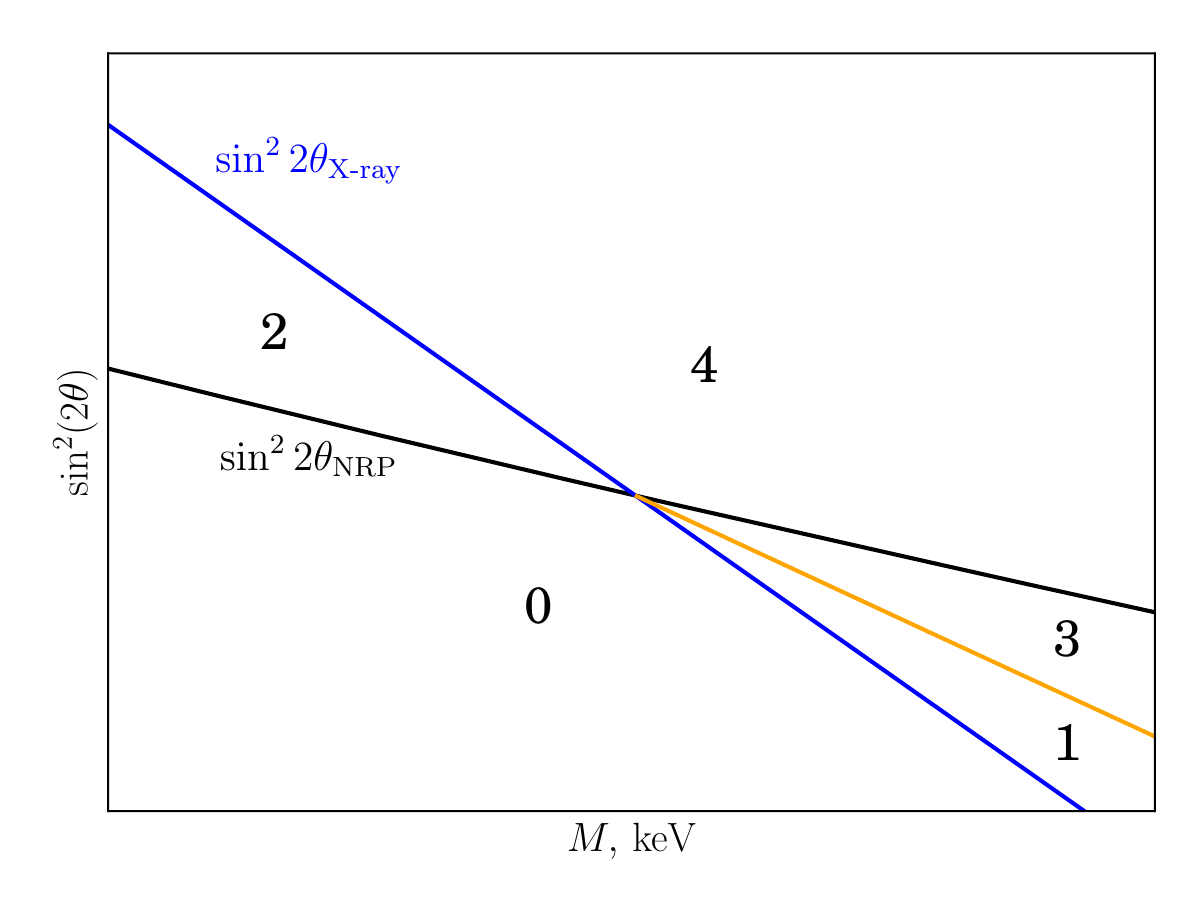}
  \caption{A sketch of astrophysical and cosmological constraints
    imposed on the sterile neutrino model.  The black line refers to
    the non-resonant production of sterile neutrinos with
    $\Omega_N=\Omega_\mathrm{DM}$, the blue line indicates the X-ray
    constraint $\theta_\text{X-ray}$ for
    $\Omega_N=\Omega_\mathrm{DM}$, the yellow line shows the
    geometrical mean of the two constraints.  Regions 0 and 1 do not
    contradict the astrophysical constraints for any $T_c$, in regions
    2, 3, 4 suppression of the sterile neutrino production is
    required, $T_c\lesssim T_\text{max}$.}
  \label{fig:generic}
\end{figure}

With a sketch in figure~\ref{fig:generic} we illustrate the role of
each constraint in different areas in the plane of sterile neutrino
mass and mixing angle.  The black line corresponds to the mixing
angle, $\sin^22\theta_\text{NRP}$, leading to the generation in
oscillations of DM amount of sterile neutrinos,
$\Omega_N=\Omega_\mathrm{DM}$ (non-resonant production).  In the
scenarios described in present work this line corresponds to
$T_c\to\infty$.  The blue line is the X-ray constraint for sterile
neutrinos $\theta_\text{X-ray}$, assuming
$\Omega_N=\Omega_\mathrm{DM}$.  Then, in the region 0 the sterile
neutrino composes only a part of DM, and is not constrained by X-ray
observations, even without introduction of the hidden sector with
phase transition and low critical temperature $T_c$.  The same is true
in the region 1, which lies below the line
\begin{equation}
  \label{geom-angle}
  \theta_\text{max\,X-ray}^2(M) =
    \sqrt{\theta^2_\text{X-ray}(M) \theta^2_\text{NRP}(M)}.
\end{equation}
Really, the amount of neutrino produced here by the non-resonant
oscillations is small, and the X-ray bound (\ref{eq:Xray}) is
satisfied.

In the region 2 it is possible to suppress the amount of produced
sterile neutrinos by reducing $T_c$ (see eq.~\eqref{eq:Omegalate}),
and make $\Omega_N=\Omega_\text{DM}$.  However, the DM composed out of
sterile neutrino produced in this way strongly violates the structure
formation bounds. Therefore, only some part of DM can be generated
here.  A rough constraint on this part can be obtained from
\cite{Diamanti:2017xfo}, but we do not present a detailed analysis
here (in our case this constraint is always satisfied once the sterile
neutrino fraction does not exceed roughly one third).

In the regions 3 and 4 the overall allowed abundance of the sterile
neutrino is determined solely by the X-ray bound (\ref{eq:Xray}).  The
respective $T_c$ is obtained from combination of (\ref{eq:Xray}) and
(\ref{eq:Omegalate}) (or, in case of the model with phase transition,
of the sum of (\ref{eq:Omegalate}) and (\ref{Omega-initial})).

For moderate values of $\theta$, which are interesting for the direct
laboratory searches, $T_c$ should be quite low, in order to
significantly suppress the sterile neutrino production.  In the
scenario with phase transition (section~\ref{sec:PhaseTransition}) the
suppression can not be arbitrary due to the production
(\ref{eq:fSfAinitial}) of neutrino above $T_c$, thus leading to the
maximal allowed mixing angle.  For the scenario of
section~\ref{sec:FrozenScalar} this limit is practically absent, as
far as $T_c\sim T_\text{osc}\ll M_N$ are allowed.

\begin{figure}[!htb]
  \centerline{%
    \includegraphics[width=0.55\textwidth]{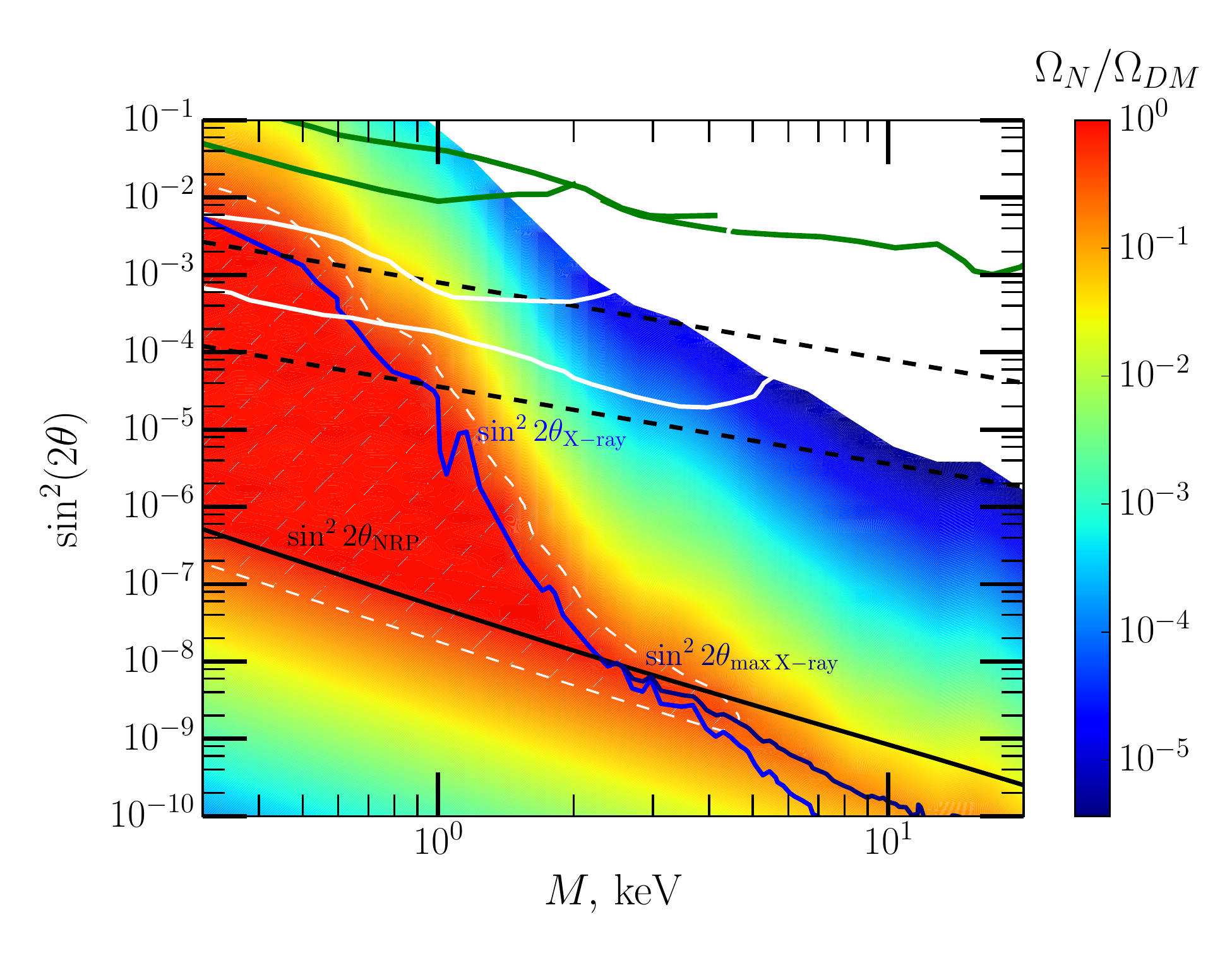}%
    \hskip -0.02\textwidth
    \includegraphics[width=0.55\textwidth]{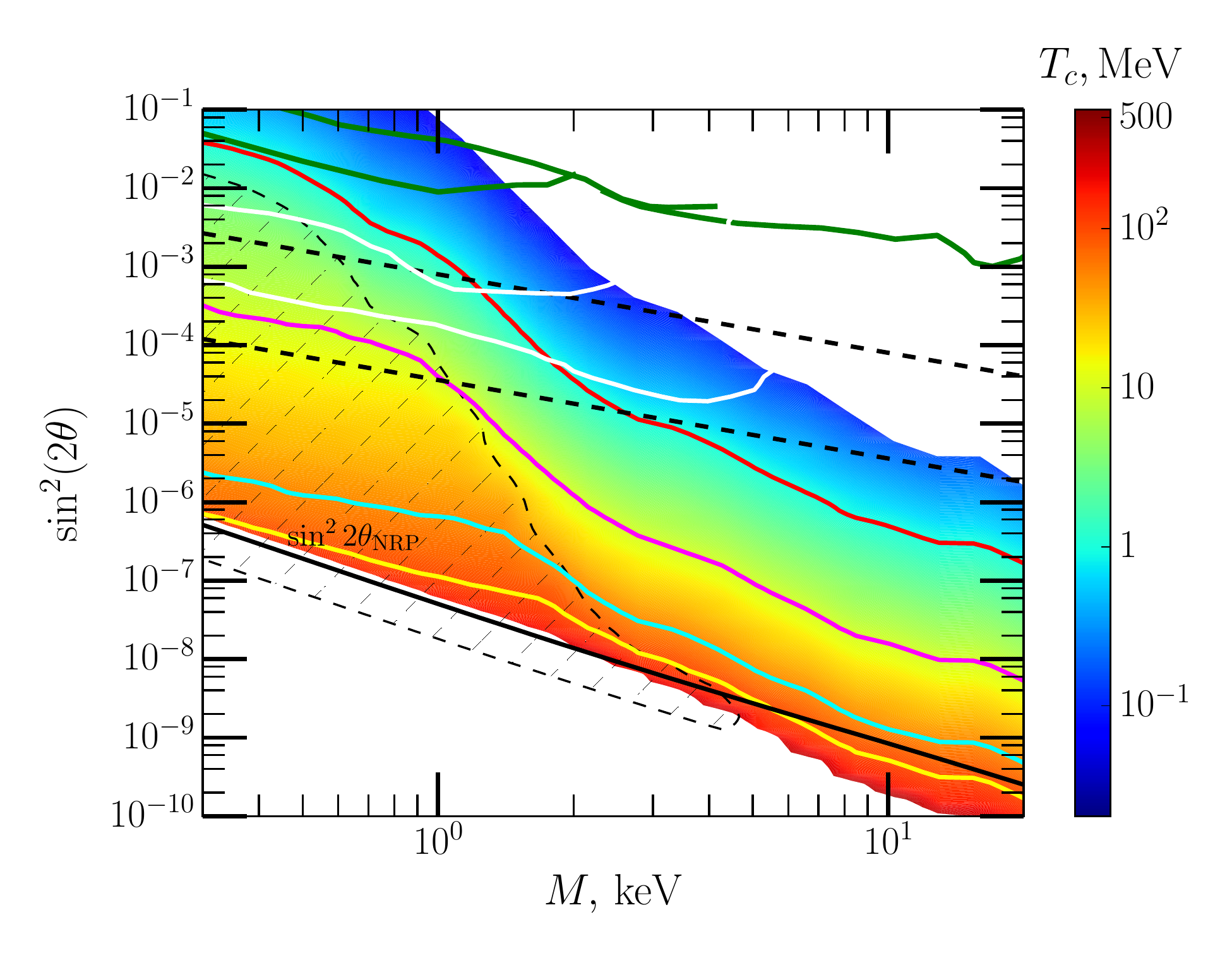}%
  }
  \vspace{-0.5cm}
  \caption{The numerical results for the model of
    section~\ref{sec:PhaseTransition}. Maximal fraction of sterile
    neutrinos in total Dark Matter $\Omega_N/\Omega_\mathrm{DM}$ that
    agrees with the X-ray bound (\ref{eq:Xray}) and does not exceed
    the DM abundance (left panel) and corresponding phase
    transition temperatures $T_c$ (right panel) are indicated by
    colour.  Green curves on top of the plots give the direct
    constraints from particle physics experiments
    \cite{Hiddemann:1995ce,Holzschuh:1999vy,Abdurashitov:2017kka}.
    Two solid white lines are projected sensitivity of Troitsk
    $\nu$-mass after two stages of
    upgrade~\cite{Abdurashitov:2015jha}.  Inclined dashed black lines
    indicate the reference seesaw values \eqref{seesaw} for active
    neutrino masses between $0.2\eV$ and $0.009\eV$. Black solid curve
    on both plots corresponds to non-resonant production of
    $\Omega_N=\Omega_\mathrm{DM}$.  In the region diagonally dashed by
    white (left) and black (right) lines sterile neutrinos can
    contribute more than $30\%$ of whole DM abundance on the left and
    right panels, respectively, so stronger suppression is required to
    satisfy the structure formation constraints.  \emph{Left panel:}
    Blue solid curve shows X-ray constraint $\theta_\text{X-ray}$
    for $\Omega_N=\Omega_\mathrm{DM}$.  Dark blue solid line is
    \eqref{geom-angle}.  \emph{Right panel:} yellow, cyan, magenta and
    red curves refer to the following temperatures of phase
    transition $T_c=100,\,50,\,10$ and $1\MeV$. White region in the
    lower left corner denotes the area where the phase transition is
    not required ($T_c\to\infty$).}
  \label{fig:phaseTr}
\end{figure}

Figure~\ref{fig:phaseTr} gives the numerical results obtained using
the full numerical solution of (\ref{eq:BoltzmanRate}) and
(\ref{eq:fSfAinitial}) in the scenario of
section~\ref{sec:PhaseTransition}.  The plots show the \emph{maximal}
possible sterile neutrino abundance $\Omega_N/\Omega_\text{DM}$ and
\emph{maximal} possible $T_c$ that satisfies the total abundance and
X-ray constraints.  Below we refer to the particular parts of plots in
Figure~\ref{fig:phaseTr} by making use of the names introduced in
figure~\ref{fig:generic}.

In the regions 0 and 1 no $T_c$ is required, and $\Omega_N$ just
corresponds to non-resonant production.  In regions 3 and 4 the
maximal abundance saturates the bound (\ref{eq:Xray}).  The
upper-right white region corresponds to the area, where required
suppression can not be achieved in the model of
section~\ref{sec:PhaseTransition}.

In the region 2 of the plot the X-ray bound allows for
$\Omega_N=\Omega_\text{DM}$.  However, as mentioned above, this would
strongly violate the structure formation constraints.  Even rather
conservative constraints give for the case of non-resonantly produced
DM the bounds of $m_\text{NRP}>8\keV$ from Lyman-$\alpha$
\cite{Boyarsky:2008xj} and of $m_\text{NRP}>5.7\keV$
\cite{Gorbunov:2008ka} from phase space density.  For the sterile
neutrino with distribution (\ref{eq:fSfAapprox}) we can rescale the
mass as
\begin{equation}
  \label{mnrp}
  m_\text{NRP} = \frac{\lAngle p \rAngle|_{f_A}}{\lAngle p \rAngle|_{f_N}} M
  = \frac{3.1}{4.1} M,
\end{equation}
obtaining the constraints of $M>11\keV$ and $8\keV$, respectively,
which exclude the whole region.\footnote{Strictly speaking, the
  distribution of sterile neutrinos for the model of
  section~\ref{sec:PhaseTransition} has also the colder component
  (\ref{eq:fSfAinitial}) with $\lAngle p\rAngle=1.2T$, but it is not
  sufficient to relax the structure formation constraints
  significantly enough.}  It is hard to precisely estimate what
fraction of Warm DM is still allowed, so we do not take the structure
formation constraint into account in the plot, but dashed the region
where they are important (note, that it is slightly larger than the
area 2 of the sketch in figure~\ref{fig:generic}, as far as it
corresponds to the region where the amount of the produced WDM
component is above 30\%).

Note, that in the model with the feebly interacting scalar of
section~\ref{sec:FrozenScalar} there may be nonperturbative generation
of super-cool sterile neutrinos by the oscillating background.  This
may lead to sterile neutrino Cold DM, thus evading the structure
formation constraints in region 2.  We leave the investigation of this
possibility for future work.

It is clearly seen from figure~\ref{fig:phaseTr}, that in the model of
section~\ref{sec:PhaseTransition} with allowed
$T_c>T_{c,\mathrm{min}}$~\eqref{eq:Tcmin} it is possible to reach the
area of the direct ground-based constraints (green solid lines in
figure~\ref{fig:phaseTr}).  In the model of
section~\ref{sec:FrozenScalar} even larger values of mixing angle are
compatible with the cosmological constraints.

Note in passing, that in our analysis sterile neutrino mixes with the
electron active neutrino only.  However, the upper bound of available
region (the upper edges of colour areas in figure~\ref{fig:phaseTr})
remains valid for the case of mixing with muon and tau species.  The
reason is that at low temperatures (below electron decoupling) there
is no difference in production rate of sterile neutrinos from mixing
with any active species. The X-ray limits are the same for mixing with
all the three active neutrinos.

Finally, the model of section~\ref{sec:PhaseTransition} with phase
transition at some moment in the past leads to significant extension
of available region in the plane $(M,\theta^2)$. In the model of
section~\ref{sec:FrozenScalar} with feebly interacting scalar, the
sterile neutrino production in the early Universe can be fully
suppressed making X-ray limits irrelevant.  Then all the region below
the direct limits is allowed. Moreover, the presently developing
projects, e.g.~\cite{Abdurashitov:2015jha}, can directly probe the
mixings~\eqref{seesaw} consistent with minimal implication of the
seesaw type-I mechanism of generation of active neutrino masses.

\section{Conclusions}
\label{sec:conclusions}

To summarise the results in one sentence, we show how to suppress the
sterile neutrino production in the early Universe to the level which
makes the direct laboratory searches competitive to the investigations
with the X-ray satellites of the radiative decays of the cosmic
sterile neutrino background.

Mechanisms suggested in the paper the sterile neutrino production can
be suppressed to the level where even present direct limits become the
strongest ones. This shows the importance of the direct searches, in
particular in exploring the see-saw type I mechanism of active
neutrino mass generation.  On the contrary, the astrophysical indirect
searches are subject to model-dependent uncertainties.

The possible scenarios are not limited by the two outlined, but the
common statement is that the change of the sterile neutrino mass
should happen at $T_c<T_{\mathrm{max}}$, and that requires hidden
sector with new physics at low energy scale.

The authors are indebted to Igor Tkachev for valuable discussions and
help with depicting the very recent limits~\cite{Abdurashitov:2017kka}
of Troitsk $\nu$-mass experiment. The works is supported by the RSF
grant 17-12-01547.

\bibliographystyle{JCAP-hyper}
\bibliography{variablemass}

\end{document}